\documentclass[twocolumn,showpacs,preprintnumbers,amsmath,amssymb]{revtex4}
\usepackage{graphicx}
\usepackage{dcolumn}
\usepackage{bm}

\begin{document}

\title{Spin-dependent neutrino-induced nucleon knockout}
\author{N.~Jachowicz}
\email{natalie.jachowicz@UGent.be}
\author{K.~Vantournhout, J. Ryckebusch, and K.~Heyde}
\affiliation{Department of Subatomic and Radiation Physics,\\ Ghent University, \\Proeftuinstraat 86, \\ B-9000 Gent, Belgium.} 
\date{\today}

\pacs{25.30.Pt,13.15.+g,25.30.-c,13.88.+e,24.70.+s,26.50.+x}

\begin{abstract}

We study neutrino-induced nucleon knockout off atomic nuclei and examine the polarization properties of the ejectile. A detailed study of the spin dependence of the outgoing nucleon is presented.  
The numerical results are  derived within a non-relativistic plane-wave impulse-approximation approach. Our calculations reveal large polarization asymmetries, and clear dissimilarities between neutrino- and antineutrino-induced reactions. They reflect the fact that neutrino-induced nucleon knockout is dominated by the transverse axial current and gains its major contributions from forward nucleon emission and backward lepton scattering.

\end{abstract}

\pacs{25.30.Pt, 26.50.+x, 14.60.Lm}
\maketitle

\section{Introduction}\label{intro}

 In numerous processes governed by the weak Hamiltonian, its  parity violating character becomes manifest.
The interaction couples to the left- and righthanded components of the particles involved in weak processes in a different way.
The equal weight of vector and axialvector contributions to the  lepton current implies that the weak interaction couples exclusively to lefthanded leptons and to righthanded antileptons.  
As neutrinos only interact by means of the weak Hamiltonian, it is in their behavior that the parity violation becomes most obvious.  The maximal parity violation in the lepton sector assures that weakly interacting neutrinos are lefthanded, antineutrinos are righthanded.

Similar mechanisms control the quark couplings. In the $SU(2)\otimes U(1)_Y$ description of the electroweak interaction, first-generation quarks are introduced as a lefthanded doublet and two righthanded singlets. 
In charged-current reactions the weak interaction couples only to the lefthanded quark doublet. For neutral-current interactions, the mixture of a weak neutral boson and an electromagnetic $U(1)_{EM}$ gauge boson  to the weak neutral $Z^{0}$-boson, results in a coupling to the righthanded quark components.  
 
Despite the prominent presence of parity-breaking axial contributions in weak-interaction processes, little attention has been paid to induced polarizations
in neutrino-nucleus scattering processes. In electron scattering on the other hand, polarization phenomena attracted a lot of attention and have  been the subject of numerous experimental and theoretical studies \cite{jan,lava,poltrans,strauch1, strauch2}.
In recent years, a large number of theoretical studies on neutrino scattering was carried out
including Fermi gas, random phase approximation, and shell model calculations \cite{Ed,kim,kol1,kol2,albe,sing,ikke0,fuller,town,vol,edw,ikke1,ikke2,ikke3,volpenieuw,edlan,crist,meu,nieves,kub1,kub2,kub3}.  Reactions of experimental \cite{kol2,town,vol,ikke1,crist,meu,nieves} as well as astrophysical \cite{edw,fuller,ikke2,ikke3,volpenieuw,Heger} interest were examined.
The spin dependence of neutrino-induced nucleon knockout was noted in \cite{letter}.  In the present paper, we extend this work to  a systematic study of the ejectile's  spin polarization in neutrino scattering off atomic nuclei. Expressions for the spin dependence of  neutrino-induced nucleon knockout cross sections                    are derived. We examine the mechanisms underlying the observed asymmetries.
The results are  illustrated in a non-relativistic plane-wave impulse approximation approach. We present cross sections  and estimate the impact of polarization asymmetries on the neutrino-induced nucleon knockout processes. 

\section{Neutrino-nucleus scattering cross sections}\label{formalism}

\begin{figure}
\vspace*{6cm}
\special{hscale=48 vscale=48 hsize=1500 vsize=600
         hoffset=0 voffset=15 angle=-0 psfile="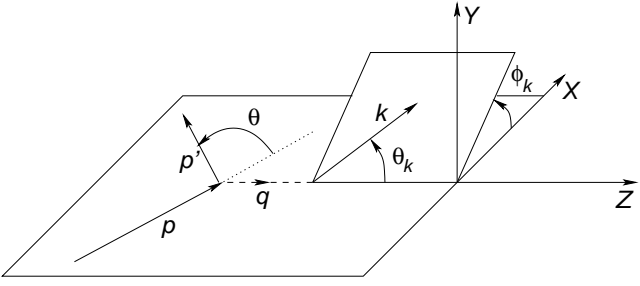"}
\caption{Kinematics of the described neutrino scattering reactions. The angle between the scattering and the reaction plane is denoted by $\phi_k$, $\theta_k$ is the direction of the outgoing lepton relative to the momentum exchange $\vec{q}$. }
\label{kin}
\end{figure}

The kinematics of the neutral-current reactions 
\begin{eqnarray}
\nu(\varepsilon_i,\vec{p})+A &\rightarrow& (A-1)+\vec{N}(E_{N},\vec{k})
+\nu'(\varepsilon_f,\vec{p'})\\
\overline{\nu}(\varepsilon_i,\vec{p})+A &\rightarrow& (A-1)+\vec{N}(E_{N},\vec{k})
+\overline{\nu}'(\varepsilon_f,\vec{p'}),
\end{eqnarray}
we are considering are presented in Fig.~\ref{kin}.  A neutrino with incoming momentum $\vec{p}$ and energy $\varepsilon_i$ exchanges a four momentum $q=p-p'$ with a nucleon in the nucleus.  The outgoing nucleon's momentum is $\vec{k}$, the lepton leaves  in a direction $\theta$ relative to the incoming  beam.
As in neutral-current neutrino-scattering the outgoing lepton cannot be detected, 
the momentum exchange remains unknown and the missing momentum cannot be reconstructed.
Therefore the direction of the outgoing nucleon $\theta_{N}$ has to be  defined as the angle between $\vec{k}$ and the direction of the incident neutrino $\vec{p}$ . As a consequence, polarization studies necessarily have to focus on the longitudinal spin component i.e.~the component of the nucleon's spin along the direction of its outgoing momentum $\vec{k}$.

Under these conditions, the differential cross section can be written as
\begin{eqnarray}
\lefteqn{\frac{d^3\sigma}{d\Omega_{N}dE_{N}} }\nonumber
\\&=&(2\pi)^4  E_{N}k
\overline{\sum_{f,i}}
\left|\langle f\left|\widehat{H}_W\right|i\rangle\right|^2 \,\delta(E_i-E_f),\label{cs}
\end{eqnarray}
where $\widehat{H}_W$ denotes the weak interaction Hamiltonian, 
 which factorizes in a lepton  $\hat{\jmath}_{\mu}(\vec{x})$ and a hadron part $\hat{J}^{\mu}(\vec{x})$ 
\begin{eqnarray}
\widehat{H}_W&=&\frac{G_F}{\sqrt{2}}\int\hat{\jmath}_{\mu}(\vec{x})\hat{J}^{\mu}(\vec{x})d^3x
\\&=&\frac{G_F}{\sqrt{2}}\int e^{i\vec{q}\cdot\vec{x}}\hat{l}_{\mu}\,\hat{J}^{\mu}(\vec{x})d^3x,
\end{eqnarray}
allowing one to write the transition density as
\begin{eqnarray}
M_{fi}&=&\langle f\left|\widehat{H}_W\right|i\rangle\nonumber\\
&=&\frac{G_F}{\sqrt{2}}\int d^3 x\: \langle f_l\left|\hat{\jmath}_{\mu}\right| i_l\rangle
\langle f_h\left|\hat{J}^{\mu}\right| i_h\rangle\\
&=&\frac{G_F}{\sqrt{2}}l_{\mu} \int d^3 x\:
\langle f_h\left|e^{i\vec{q}\cdot\vec{x}}\hat{J}^{\mu}\right| i_h\rangle, \label{eq7}
\end{eqnarray}
with $|i_h\rangle$, $|i_l\rangle$, $|f_h\rangle$, $|f_l\rangle$ denoting the initial and final hadron and lepton states,
\begin{equation}
\hat{\jmath}_{\mu}=e^{i\vec{q}\cdot\vec{x}}l_{\mu},
\end{equation}
and $l_{\mu}$ the lepton current
\begin{eqnarray}
l_{\mu}=\frac{1}{2(2\pi)^3\sqrt{\varepsilon_i\varepsilon_f}}\overline{u}^{(\nu')}(p',s_f)\gamma_{\mu}(1-\gamma_5)u^{(\nu)}(p,s_i),
\end{eqnarray}
for lefthanded neutrinos and 
\begin{equation}
l_{\mu}=\frac{1}{2(2\pi)^3\sqrt{\varepsilon_i\varepsilon_f}}\overline{u}^{(\overline{\nu}')}(p',s_f)\gamma_{\mu}(1+\gamma_5)u^{(\overline{\nu})}(p,s_i),
\end{equation}
corresponding  to the righthanded helicity for antineutrinos.

The initial hadron state corresponds to the ground state of the target nucleus, the nuclear final state is determined by the quantum numbers of the outgoing nucleon and the residual nucleus
\begin{equation}
\langle f_h|\hat{J}^{\mu}|i_h\rangle=\langle\Psi_f;k,\frac{1}{2}r|\hat{J}^{\mu}|\Psi_i\rangle,
\end{equation}
with $r$ denoting the spin projection of the ejectile.
In the impulse approximation the nucleons are treated as independent particles, and the hadron current matrix elements as a sum of one-body contributions.  The different contributions to the non-relativistic limit of the hadron current are obtained as
\begin{subequations}
\begin{align}
\begin{split}
\hat{J}^0_V(\vec{x})=& \sum_{i=1}^A G_E^i(q_{\mu}^2)\;\;\delta(\vec{x}-\vec{x}_i),
\end{split}\\
\begin{split}
\hat{J}^0_A(\vec{x})=&\sum_{i=1}^A \frac{G_A^i(q_{\mu}^2)}{2m_i}\;\hat{\vec{\sigma}}_i\cdot(\hat{\vec{k}}'_i+\hat{\vec{k}}_i)\;\;\delta(\vec{x}-\vec{x}_i),
\end{split}
\\
\begin{split}
\hat{\vec{J}}^c_V(\vec{x})=&\sum_{i=1}^A\frac{G_E^i(q_{\mu}^2)}{2m_i}\;(\hat{\vec{k}}'_i+\hat{\vec{k}}_i)\;\;\delta(\vec{x}-\vec{x}_i),
\end{split}
\\
\begin{split}
\hat{\vec{J}}^m_V(\vec{x})=&\sum_{i=1}^A \frac{G_M^i(q_{\mu}^2)}{2m_i}\;\hat{\vec{\sigma}}_i\times\hat{\vec{q}}\;\;\delta(\vec{x}-\vec{x}_i),
\end{split}
\\
\begin{split}
\hat{\vec{J}}_A(\vec{x})=&\sum_{i=1}^AG_A^i(q_{\mu}^2)\;\hat{\vec{\sigma}}_i\;\;\delta(\vec{x}-\vec{x}_i),
\end{split}\end{align}\end{subequations}
with $\hat{\vec{k}}'=i\stackrel{\leftarrow}{\nabla}$ and $\hat{\vec{k}}=-i\stackrel{\rightarrow}{\nabla}$ the momentum operators working on the initial and final hadron state respectively. Denoting
\begin{equation}
h^{\mu}=\int d^3x \;\;e^{i\vec{q}\cdot\vec{x}}\langle f_h|\hat{J}^{\mu}(\vec{x})|i_h\rangle,
\end{equation}
the transitions are determined by an expression of the type
\begin{equation}\label{f1}
\left|\langle f\left|\widehat{H}_W\right|i\rangle\right|^2= \frac{G_F^2}{2} l_\mu {l_\nu}^* h^\mu h^{\nu\;*}.
\end{equation}
The lepton current contains the  projection operator $\frac{(1-\gamma_5)}{2}$ for neutrinos and its righthanded counterpart $\frac{(1+\gamma_5)}{2}$ for antineutrinos.  This factor reflects the vector-axialvector structure of the weak interaction Hamiltonian and projects the desired helicity for the neutrinos. Summing the incoming and outgoing lepton spins $s$ and $s'$, the leptonic part of Eq.~(\ref{f1}) can be rewritten as
\begin{equation}\label{f4}
\frac{(2\pi)^6}{2}\sum_{ss'}l_\mu {l_\nu}^*=\frac{p_\nu p'_\mu+p_\mu p'_\nu-pp'g_{\mu\nu}\pm ip^\alpha p'^\beta \epsilon_{\alpha\nu\beta\mu}}{\varepsilon_i\varepsilon_f},
\end{equation}
with $p_{\mu}$ and $p_{\mu}'$ the incoming and outgoing lepton momentum, $\varepsilon_i$ and $\varepsilon_f$ the corresponding energies, and $\epsilon_{\mu\nu\rho\sigma}$ the four-dimensional Levi-Civita symbol ($\epsilon_{0123}$ and even permutations equal -1, odd permutations equal +1).
After inserting Eq.~(\ref{f4}) into (\ref{eq7}) , one obtains
\begin{widetext}
\begin{eqnarray}
\sum_{ss'}|M_{fi}|^2&=&\frac{G_F^2}{(2\pi)^6\varepsilon_i\varepsilon_f}
(p_{\nu} p'_{\mu}+p_{\mu} p'_{\nu}-pp'g_{\mu\nu}\pm i p^{\alpha} p'^{\beta} \varepsilon_{\alpha\nu\beta\mu})(h^{\mu}h^{\nu *})\label{f7}\\
&=&{\frac{G_F^2}{(2\pi)^6\varepsilon_i\varepsilon_f}}
\left[(p_0 h_0-\vec{p}\cdot\vec{h})(p'_0{h_0}^*-\vec{p}'\cdot{\vec{h}}^*)+
(p'_0 h_0-\vec{p}'\cdot\vec{h})(p_0{h_0}^*-\vec{p}\cdot{\vec{h}}^*)\right.\nonumber\\&&
\hspace*{2cm}-\left.(p_0p'_0-\vec{p}\cdot\vec{p}')(h_0{h_0}^*-\vec{h}\cdot{\vec{h}}^*)
\pm i\varepsilon_{\alpha\nu\beta\mu}p^{\alpha}p'^{\beta}h^{\mu}h^{\nu *}\right],
\end{eqnarray}
and the differential cross section becomes
\begin{eqnarray}
\frac{d^3\sigma}{d\Omega_{N}dE_{N}}&=&\frac{G_F^2}{(2\pi)^2}\;\;
\overline{\sum}_{i,f}\,\int_0^{\varepsilon_i-S_N}\varepsilon_f^2 d\varepsilon_f\;\int_{4\pi}d\Omega_{\nu}\,\frac{E_{N}k}{\varepsilon_i\varepsilon_f}\nonumber\\
&&\times\left\{[p_0 h_0-\!\sum_{\lambda=\pm 1,z} (-1)^\lambda p_\lambda h_{-\lambda}][p'_0
  {h_0}^*-\!\sum_{\lambda=\pm 1,z} p'_\lambda{{h}_{\lambda}}^*]
+[p'_0 h_0-\!\sum_{\lambda=\pm 1,z}(-1)^\lambda {p}'_\lambda h_{-\lambda}][p_0
  {h_0}^*-\!\sum_{\lambda=\pm 1,z} p_\lambda{{h}_{\lambda}}^*]\right.\nonumber\\
 & &-[p_0p'_0-\vec{p}\cdot\vec{p}'][h_0{h_0}^*-\!\sum_{\lambda=\pm 1,z} h_\lambda{{h}_{\lambda}}^*]
\pm\left.
\begin{vmatrix}
p_0 & p_+ & p_- & - p_z \\
p'_0 & p'_+ & p'_- & - p'_z \\
h_0 & h_+ & h_- & -h_z \\
{h_0}^* & -{h_-}^* & -{h_+}^* & -{h_z}^*
\end{vmatrix}\right\}\,\delta(E_i-E_f),\label{8f8}
\end{eqnarray}
\end{widetext}
where $S_N$ denotes the particle emission threshold. Recoil effects were neglected.
In the following, the outgoing nucleon is described by a plane wave, and off-shell effects and final-state interactions are neglected. In the non-relativistic limit the hadronic matrix elements then read 
\begin{widetext}
\begin{subequations}
\label{hadron}
\begin{align}
\begin{split}
h^0_V (\Omega_m)=&\sum_{m_lm_s}\; \langle l\,{m_l},{\frac{1}{2}}\,{m_s}|{j}{m}\rangle \;G_E(q_{\mu}^2)\;\delta_{r,m_s}\;F_{nlm_lj}(\Omega_m),\label{ha}\end{split}\\\begin{split}
h^0 _A(\Omega_m)=&
 \sum_{\stackrel{m_lm_s}{\lambda}}\;\langle l\,{m_l},{\frac{1}{2}}\,{m_s}|{j}{m}\rangle\; G_A(q_{\mu}^2)\;(-1)^{\lambda}\,\sqrt{3}\frac{(2\vec{k}-\vec{q})_{-\lambda}}{2m_N}\;\langle \frac{1}{2}\,{m_s},1\,{\lambda}|\frac{1}{2}\,r\rangle\;F_{nlm_lj}(\Omega_m),\label{hb}\end{split}\\\begin{split}
(\vec{h}_V)_{\stackrel{\,}{\lambda=\pm1,z}}(\Omega_m) =& \sum_{m_lm_s}\;\langle l\,{m_l},{\frac{1}{2}}\,{m_s}|{j}{m}\rangle\;\left[ G_E(q_{\mu}^2)\; \frac{(2\vec{k}-\vec{q})_\lambda}{2m_N}\;\delta_{r,m_s}
-G_M(q_{\mu}^2)\;\lambda\;\sqrt{3}\frac{\kappa}{2m_N}\;\langle \frac{1}{2}\,{m_s},1\,{\lambda}|\frac{1}{2}\,r\rangle\right]\;F_{nlm_lj}(\Omega_m),\label{hc}							     \end{split}\\\begin{split}
(\vec{h}_A)_{\stackrel{\,}{\lambda=\pm1,z}}(\Omega_m)=&\sum_{m_lm_s}\;\langle l\,{m_l},{\frac{1}{2}}\,{m_s}|{j}{m}\rangle \;G_A(q_{\mu}^2)\;\sqrt{3}\; \langle \frac{1}{2}\,{m_s},1\,{\lambda}|\frac{1}{2}\,r\rangle\;F_{nlm_lj}(\Omega_m).\end{split}\end{align}\label{hd}
\end{subequations}
\end{widetext}
In these expressions, $r$ denotes the spin projection of the outgoing nucleon along the direction of the momentum exchange $\vec{q}$, with $|\vec{q}|=\kappa$. The factor $\langle l\,{m_l},{\frac{1}{2}}\,{m_s}|{j}{m}\rangle$ is the Clebsch-Gordan coefficient coupling the quantum numbers of the bound single particle states, $\langle \frac{1}{2}\,{m_s},1\,{\lambda}|\frac{1}{2}\,r\rangle$ is
taking care of spin conservation in the interaction,  and the spatial contribution is given by
\begin{equation}
F_{nlm_lj}(\Omega_m)=\frac{4\pi i^l}{(2\pi)^{3/2}}\int\limits_0^\infty\phi_{nlj}(r)j_l(ur) r^2dr\;\;Y_l^{m_l}(\Omega_m),
\end{equation}
with $\phi_{nlj}(r)$ the bound state wave functions and $u=|\vec{p}_m|=|\vec{q}-\vec{k}|$ the missing momentum.

Thus far, spin projections $r$ were taken along the momentum transfer $\vec{q}$ chosen as $z$-axis.   As the momentum exchange and the missing momentum cannot be determined in neutral-current reactions, the only well-defined outgoing direction is the one along the outgoing nucleon's momentum.  Therefore in the following study of the polarization of the outgoing nucleon, we will focus on its longitudinal spin component $s_N^l$. Hence an additional $SU(2)$ rotation determined by ($\theta_k$, $\phi_k$) is needed to obtain 
\begin{eqnarray}
\lefteqn{\langle \Psi_f ; k, \frac{1}{2}s_N^{l}
\left|\hat{J}^{\mu}\right|\Psi_{i}\rangle}\nonumber\\
&=&\sum_{r} {D^{(\frac{1}{2})}_{r,s_N^{l}}}^*(\theta_k,\,\phi_k)\langle \Psi_f ; k, \frac{1}{2}r 
\left|\hat{J}^{\mu}\right|
\Psi_{i}
\rangle,\label{su2}
\end{eqnarray}
with $s_N^{l}$ the sought longitudinal spin projection and  $D^{(\frac{1}{2})}$ the Wigner $D$-matrix for $SU(2)$ rotations.

\section{Nucleon polarizations in neutrino scattering}

\begin{figure}
\vspace*{7cm}
\special{hscale=35 vscale=35 hsize=1500 vsize=600
         hoffset=-16 voffset=200 angle=-90 psfile="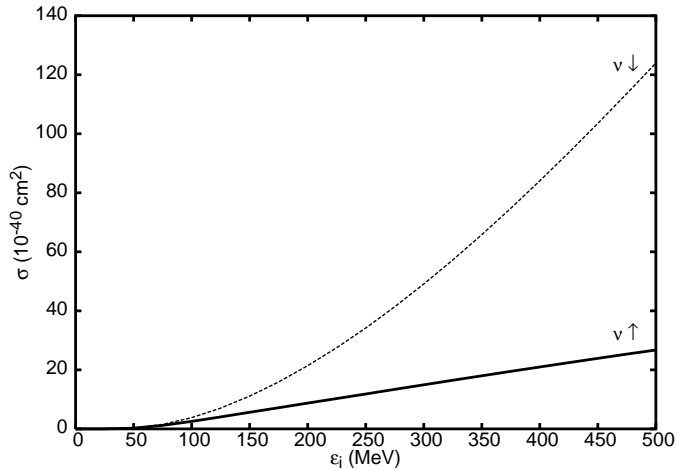"}
\caption{Cross section for neutral-current neutrino-induced nucleon knockout  from $^{16}$O, separated in its different polarization contributions ($s_N^l=+1$ : full line, $s_N^l=-1$ : dashed) for the outgoing nucleon, as a function of the incoming neutrino energy $\varepsilon_i$.}
\label{ud}
\end{figure}

\begin{figure*}
\vspace*{7cm}
\special{hscale=35 vscale=35 hsize=1500 vsize=600
         hoffset=-8 voffset=200 angle=-90 psfile="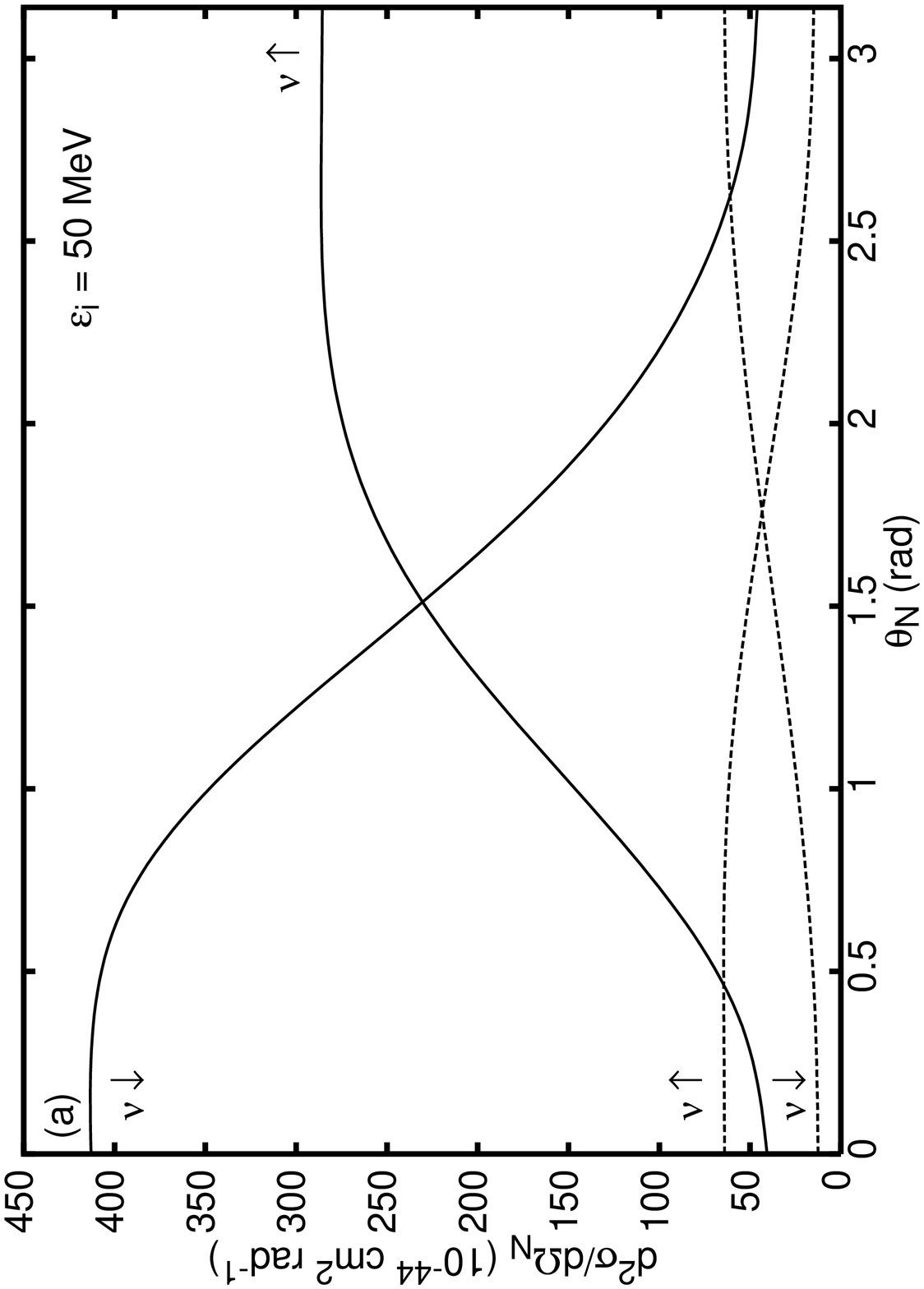"}
\special{hscale=35 vscale=35 hsize=1500 vsize=600
         hoffset=240 voffset=200 angle=-90 psfile="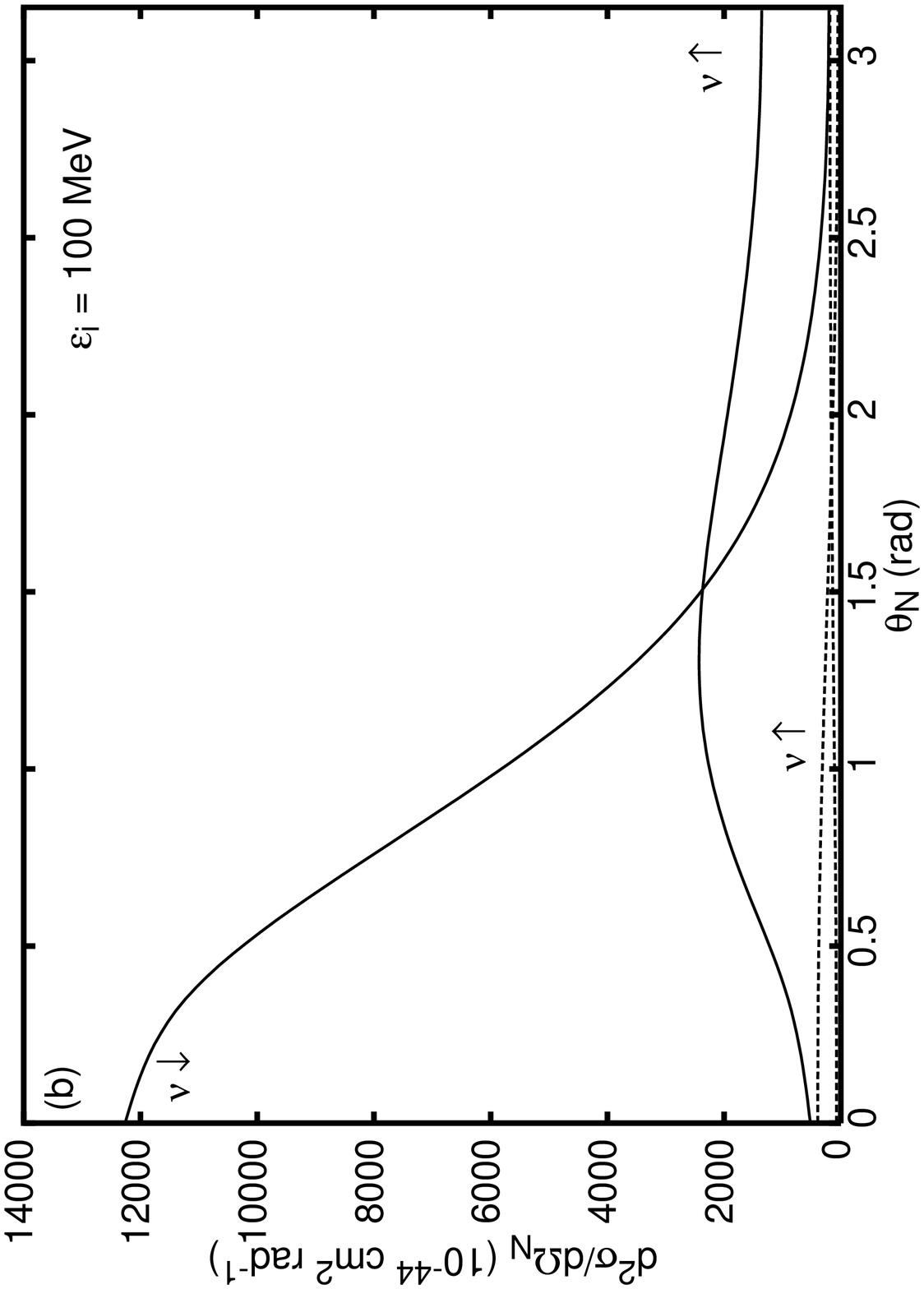"}
\caption{Total cross section  (full line) and longitudinal contribution (dashed) for  spin-up ($s_N^l=+1$) and spin-down ($s_N^l=-1$) nucleon knockout from $^{16}$O, as a function of the scattering angle of the nucleon relative to the direction of the incoming lepton $\theta_N$.  The left panel shows the results for an impinging neutrino energy of 50 MeV, in the right panel the incoming energy is 100 MeV. For 50 MeV cross sections, the difference between the total and the longitudinal results at  large scattering angles, is due to negative transverse-longitudinal interference contributions.} 
\label{long}
\end{figure*}
\begin{figure}
\vspace*{7cm}
\special{hscale=35 vscale=35 hsize=1500 vsize=600
         hoffset=-16 voffset=200 angle=-90 psfile="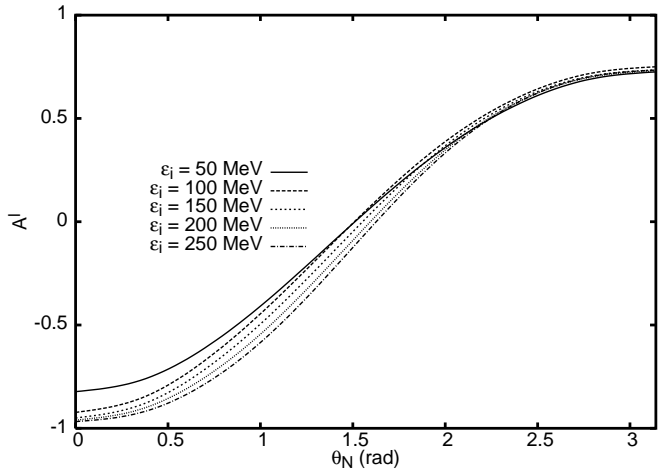"}
\caption{Longitudinal asymmetry $A^l$ in neutrino-induced nucleon knockout as a function of the ejectile's scattering angle for various values of the incoming neutrino energy : $\varepsilon_i$=50 MeV (full line), $\varepsilon_i$=100 MeV (dashed), $\varepsilon_i$=150 MeV (short-dashed), $\varepsilon_i$=200 MeV (dotted), $\varepsilon_i$=250 MeV (dash-dotted). }
\label{anl}
\end{figure}

With the expressions derived in Eqs.~(\ref{hd}), the polarization properties of the ejectile in $A(\nu,\,\nu'\:N)$ reactions  can be studied in a rather transparent way.  Our results are obtained treating the  nuclear ground state in a mean-field approach with a Woods-Saxon single-particle potential. Woods-Saxon parameters were taken from \cite{wso,wspb}.

Figure \ref{ud} displays the difference between neutrino-induced $s_N^l=+1$ ('spin-up') and $s_N^l=-1$ ('spin down') nucleon knockout from  $^{16}$O. The  
dominance of reactions with ejectile polarizations anti-parallel to their momentum is obvious.   The difference stems from the spin dependence of the hadronic matrix elements in  Eqs.~(\ref{hadron}) and grows  steadily with incoming neutrino energies.

Figure \ref{long} explains how these polarization dissimilarities arise from the different cross section contributions, for various values of the incoming neutrino-energy.
The figure presents the total and the longitudinal cross sections as a function of the scattering direction of the outgoing nucleon, relative to the incoming lepton direction.
Whereas the total cross section is governed by spin-down nucleon knockout, the longitudinal response receives its largest contribution from $s_N^l=+1$ nucleons. In addition, the figure  clearly illustrates that the cross section peaks strongly for nucleon knockout along the incoming lepton momentum ($\theta_N\approx 0$).
Both effects are due to the obvious dominance of the transverse response, which strongly favors spin-down knockout and backward lepton scattering. The inclination towards alignment of momentum transfer and ejectile momentum then causes the nucleon knockout to lign up with the incoming neutrino direction, while the spin of the ejectile is anti-parallel to its momentum.  This mechanism becomes even more  pronounced with increasing incoming neutrino energies.  The rather mild forward preference in the left panel of Figure \ref{long} becomes incontestable in the right panel where the energy of the incoming neutrino was risen to 100 MeV : enhancing the momentum transfer diminishes the  impact of the initial nucleon momentum and after integration over missing momentum a distinct forward-scattering peak remains.

Figure \ref{long} furthermore illustrates the angular dependence of the polarization asymmetries.
 For the total angular cross section, the dissimilarity in the induced nucleon polarizations is most pronounced in forward knockout.  The differences remain large over a rather broad angular range, then switching  fast to a completely opposite behavior.  As the polarization observed at $\theta_N=0$ is, even at low incoming neutrino-energies, strongly dominated by quasi-elastic scattering conditions, 
it inherits the characteristics dictated by the spin dependence of the equations in section \ref{formalism}.  Going to larger scattering angles however, the momentum transfer and the ejectile momentum diverge and the required $SU(2)$ rotation of Eq.(\ref{su2}) becomes more and more influential, eventually reversing the polarization completely at $\theta_N=\pi$.
This behavior slightly limits the importance of the spin-down nucleon knockout for low incoming energies, while the reigning transverse forward ejectile knockout always  assures the prominence of $s_N^l=-1$ in the cross sections. At higher impinging neutrino energies, backward scattering becomes rather unimportant, as does $s_N^l=+1$ ejectile knockout.  With respect to the spin properties of the ejectile, the longitudinal cross section exhibits the opposite behavior, but it is strongly suppressed. 
Summarizing, the strong selectivity of $A(\nu,\,\nu'\:N)$ reactions regarding the spin orientation of $N$, can be attributed to three effects, reinforcing each other :
first, the strong transverse dominance in the cross section, second its prominence in backward neutrino scattering going hand in hand with forward nucleon knockout, and third the strong spin dependence of the axial contribution in this transverse response.

The differences in the induced ejectile spins 
 are clearly reflected by the polarization asymmetry $A$, 
defined as the difference in ejectile polarization, normalized to the total nucleon knockout cross section~:
\begin{equation}
A^l=\frac{\sigma(s_N^l=+1)-\sigma(s_N^l=-1)}{\sigma(s_N^l=+1)+\sigma(s_N^l=-1)},
\end{equation}
where the superscript $l$ refers to the longitudinal spin components of the ejectile. Presenting the results in terms of asymmetries $A^l$ averts the need to measure absolute cross sections and provides a more solid way to quantify the relative contribution of the various ejectile spin polarization components.
Figure \ref{anl} displays the quantity $A^l$ as a function of the  direction of the outgoing nucleon.
The plot clearly shows the dissimilarities in the ejectile polarization.  For forward scattering the asymmetries are very large and negative, for backward nucleon knockout the differences become positive.
Although the energy dependence of these results seems mild, 
averaging the angular dependence and evaluating the knockout cross sections as a function of the ejectile energy yields a distinct energy dependence of $A^l$, and very large asymmetry values \cite{letter}.

\section{Neutrinos versus antineutrinos}\label{versus}

Neutrinos and antineutrinos  have opposite helicities.  Hence, their cross sections differ in the interference terms.
Defining the nuclear response by the weak structure functions
\begin{eqnarray}
R_L&=&|h_0-\frac{\omega}{\kappa}h_z|^2\\
R_{TL}&=&2{\cal R} [h_0({h_+}^*-{h_-}^*)]\nonumber\\&&\hspace*{1.9cm}-\frac{\omega}{\kappa}2{\cal R}[h_z({h_+}^*-{h_-}^*)]\\
R_{T}&=&h_+{h_+}^* + h_-{h_-}^*\\
R_{TT}&=&2{\cal R}(h_+{h_-}^*)\\
R_{TL'}&=&2{\cal R} [h_0({h_+}^*+{h_-}^*)]\nonumber\\&&\hspace*{1.9cm}-\frac{\omega}{\kappa}2{\cal R}[h_z({h_+}^*+{h_-}^*)]\\
R_{T'}&=&h_+{h_+}^*- h_-{h_-}^*,
\end{eqnarray}
the transition of Eq.~(\ref{f7}) can be rewritten as
\begin{eqnarray}
\sum_{ss'}|M_{fi}|^2&=& v_L R_L+v_{TL}R_{TL}+v_T R_T+v_{TT}R_{TT}\nonumber\\
&&+\,h\,(v_{TL'}R_{TL'}+v_{T'} R_{T'}),
\end{eqnarray}
with $v$ the corresponding lepton kinematic factors and $h$ the helicity of the incident particle : $h=-1$ for neutrinos, $h=+1$ for antineutrinos.

\begin{figure}
\vspace*{7cm}
\special{hscale=35 vscale=35 hsize=1500 vsize=600
         hoffset=-16 voffset=200 angle=-90 psfile="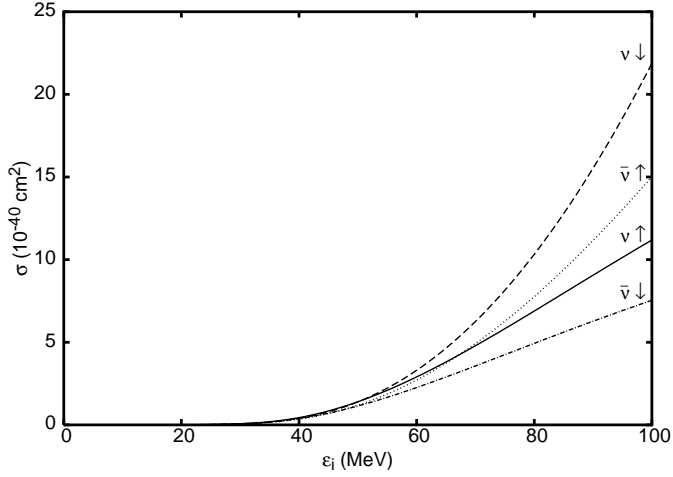"}
\caption{Neutrino- and antineutrino-induced proton-knockout cross section from $^{16}$O as a function of the incoming (anti)neutrino energy. The contributions of both ejectile polarizations are shown.}
\label{pkoa}
\end{figure}

\begin{figure}
\vspace*{7cm}
\special{hscale=35 vscale=35 hsize=1500 vsize=600
         hoffset=-16 voffset=200 angle=-90 psfile="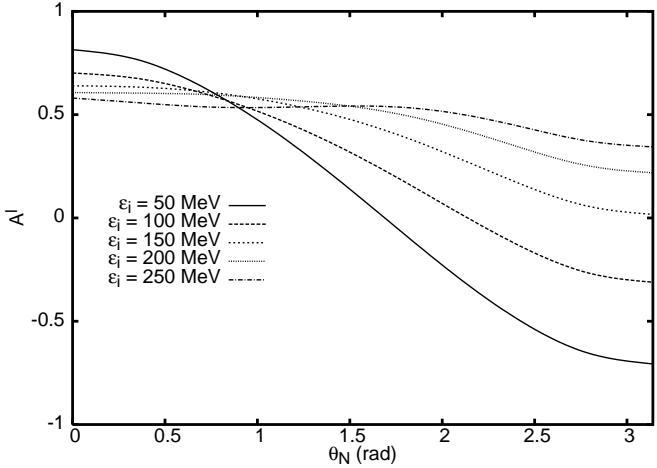"}
\caption{Longitudinal asymmetry $A^l$ in antineutrino-induced nucleon knockout as a function of the ejectile's scattering angle for different values of the incoming neutrino energy : $\varepsilon_i$=50 MeV (full line), $\varepsilon_i$=100 MeV (dashed), $\varepsilon_i$=150 MeV (shortdashed), $\varepsilon_i$=200 MeV (dotted), $\varepsilon_i$=250 MeV (dash-dotted).}
\label{aath}
\end{figure}

Except for $R_{TT}$ which is very small, the transverse terms are the most important ones.
The contribution of these terms to the cross section is   given by
\begin{eqnarray}\label{f88}
\lefteqn{l_-{l_-}^* h_+{h_+}^* + l_+{l_+}^* h_-{h_-}^*} \nonumber\\
&&=v_T(h_+{h_+}^*+ h_-{h_-}^*)\nonumber\\&&\hspace*{2.5cm} +h\, v_{T'}(h_+{h_+}^* - h_-{h_-}^*) \label{p1}\\
&&=(v_T+h v_{T'}) h_+{h_+}^* + (v_T-h v_{T'}) h_-{h_-}^*\label{p2},
\end{eqnarray}
with the symmetric and antisymmetric kinematic factors $v_T$ and $v_{T'}$
\begin{eqnarray}
v_T &=& 
\frac{2\sin^2\frac{\theta}{2}(\epsilon_i^2+\epsilon_f^2+2\sin^2\frac{\theta}{2}\epsilon_i\epsilon_f)}{(2\pi)^6(\epsilon_f^2+\epsilon_i^2-2\epsilon_i\epsilon_f\cos\theta)}, \label{s}\\
v_{T'} &=&
\frac{2\sin^2\frac{\theta}{2}(\epsilon_i+\epsilon_f)}{(2\pi)^6\sqrt{\epsilon_f^2+\epsilon_i^2-2\epsilon_i\epsilon_f\cos\theta}}.\label{a}
\end{eqnarray}
Eq.~(\ref{p1}) highlights the cross section differences between neutrino- and antineutrino-induced reactions, whereas Eq.~(\ref{p2}) governs the polarization behavior of the outgoing nucleon.   Clearly, the only difference between neutrino and antineutrino scattering is in the sign of the antisymmetric term $v_{T'}(h_+{h_+}^* - h_-{h_-}^*)$, the symmetric part contributes to both scattering reactions in the same way.  Fig. \ref{pkoa} shows these differences between neutrino- and antineutrino-induced reactions. Obviously, the  contribution $(h_+{h_+}^* - h_-{h_-}^*)$ is generally negative, resulting in antineutrino cross sections being smaller than their neutrino counterparts. The difference 
is rising steeply with the incoming lepton energy, as can  easily be explained looking at table I. 
  For  $h_+{h_+}^*$ the main axial contribution $G_A^2$ and the axial-magnetic cross term $G_AG_M\frac{\kappa}{2m_N}$ have the same sign, for $h_-{h_-}^*$ these terms have opposite signs. As  $G_AG_M$ is always negative, $h_+{h_+}^*$ is smaller than  $h_-{h_-}^*$  and the  contribution of $h_-{h_-}^*$ increases for growing energy transfers, making $|h_+{h_+}^* - h_-{h_-}^*|$ larger.

Table I
shows that the term $h_+{h_+}^*$ is dominating reactions resulting in a 'spin-up' outgoing nucleon, and the term $h_-{h_-}^*$ provides the largest contribution for 'spin-down' nucleons.
Recombining the terms in expression (\ref{p1}) to obtain Eq.~(\ref{p2}), then allows to judge the relative importance of both knockout polarizations in neutrino and antineutrino-induced reactions.
The kinematic factors $v_T$ and $v_{T'}$ are of the same order of magnitude and they have the same sign.  As a consequence,  for neutrinos the forfactor of the spin up contributions $v_T-v_{T'}$ becomes very small, resulting in a suppression of  nucleon knockout with their spin aligned with their momentum, while the $h_-{h_-}^*$ contribution is enhanced by the large factor $v_{T}+v_{T'}$.
For antineutrinos the effect is reversed.  Hence, neutrino cross sections are dominated by outgoing nucleons with their spin anti-parallel to their momentum, for antineutrino-induced reactions ejectiles with their spin aligned with the direction of their momentum are prevailing.  

Fig.~\ref{pkoa} shows  that the polarization dissimilarities are considerable. The differences indicated by Eq.(\ref{p2}) are slightly smoothed out by the spin rotations and angular averaging, but the  overall asymmetry remains.

Fig.~\ref{aath} points to clear differences in the antineutrino asymmetry $A^l$ compared to  the equivalent picture for neutrinos in figure \ref{anl}.
As is clear from the previous discussion, at forward scattering angles the asymmetry $A^l$ is positive, indicating the dominance of antineutrino-induced $s_N^l$=+1  nucleon knockout.  
The angular dependence of the asymmetry is however less pronounced than in the neutrino case, especially at larger values for the incoming lepton energy.  This is due to the influence of the energy-dependence in the cancellation effect between the
$G_A^2$ and $G_AG_M$ terms in $h_+{h_+}^*$ as they appear in Table I.

\begin{figure*}
\vspace*{12cm}
\special{hscale=38 vscale=38 hsize=1500 vsize=600
         hoffset=-8 voffset=380 angle=-90 psfile="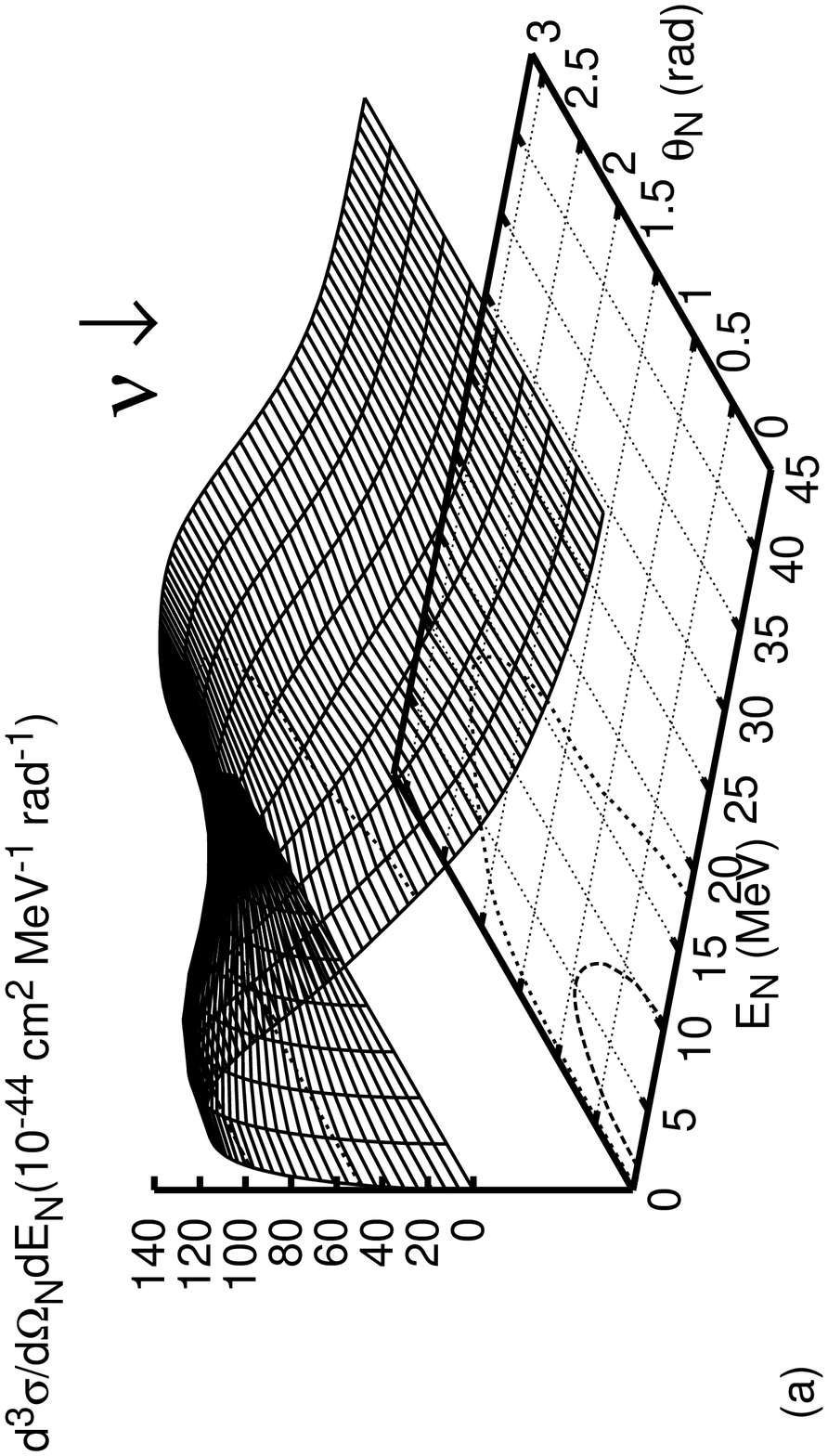"}
\special{hscale=38 vscale=38 hsize=1500 vsize=600
         hoffset=240 voffset=380 angle=-90 psfile="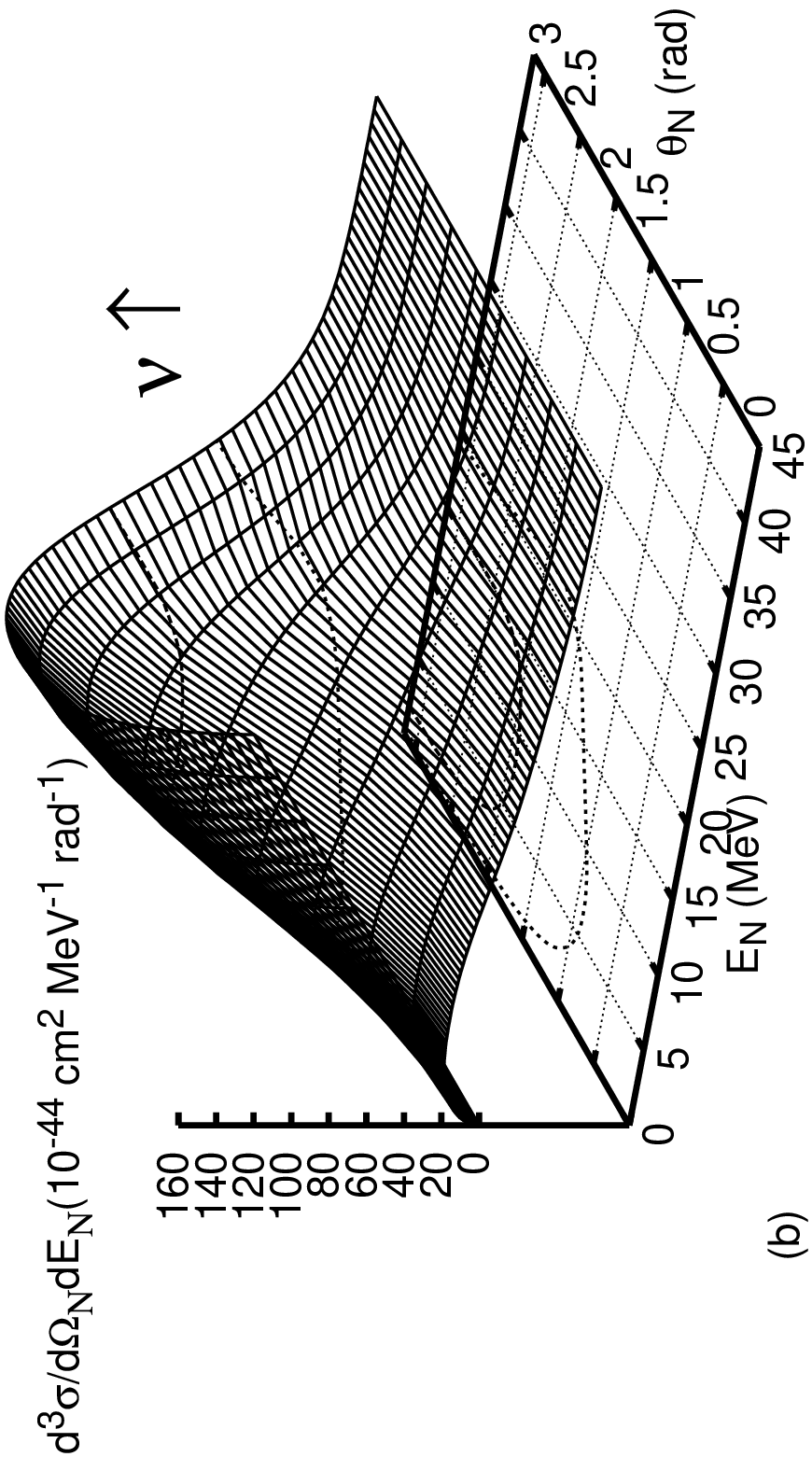"}
\special{hscale=38 vscale=38 hsize=1500 vsize=600
         hoffset=-8 voffset=200 angle=-90 psfile="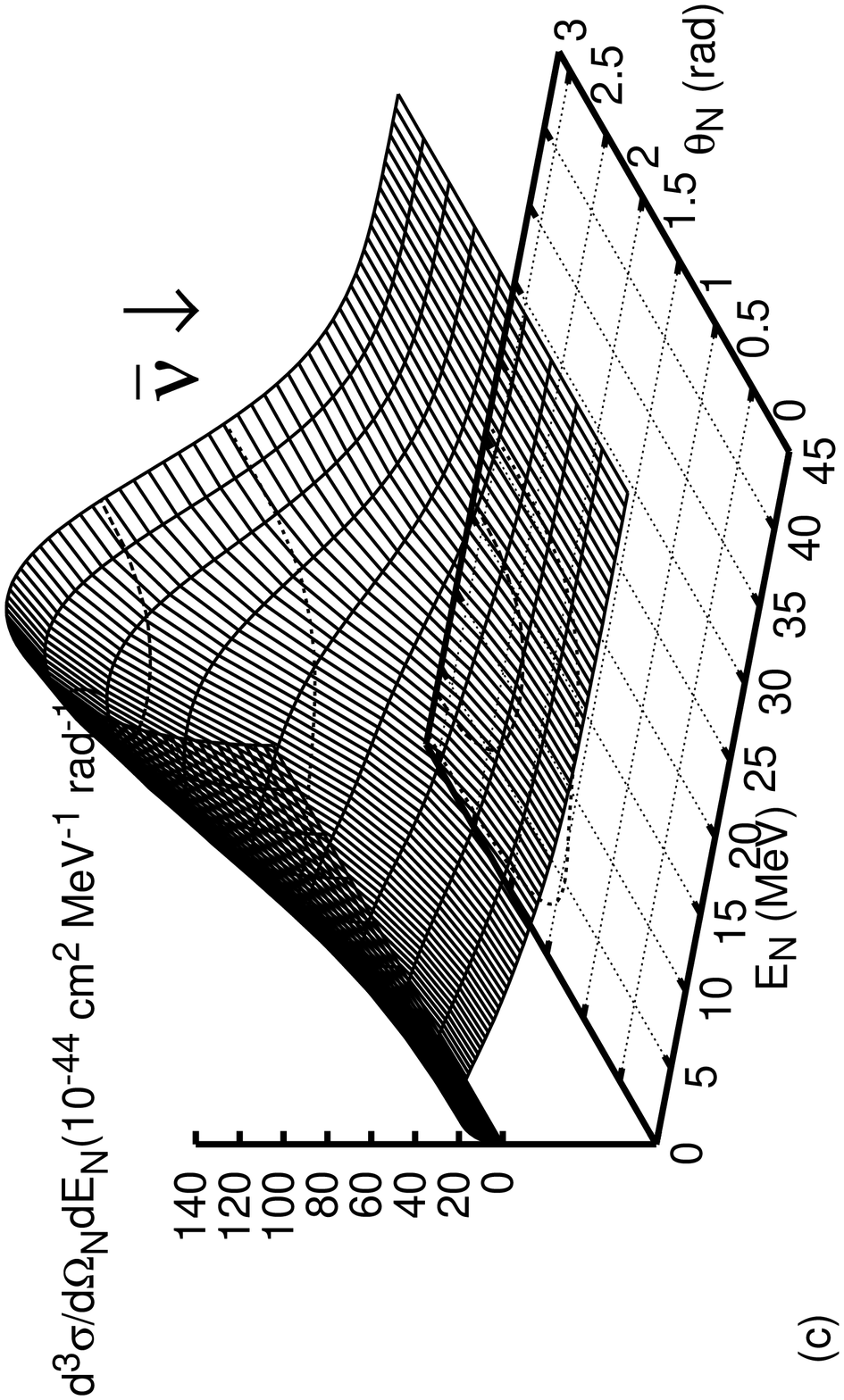"}
\special{hscale=38 vscale=38 hsize=1500 vsize=600
         hoffset=240 voffset=200 angle=-90 psfile="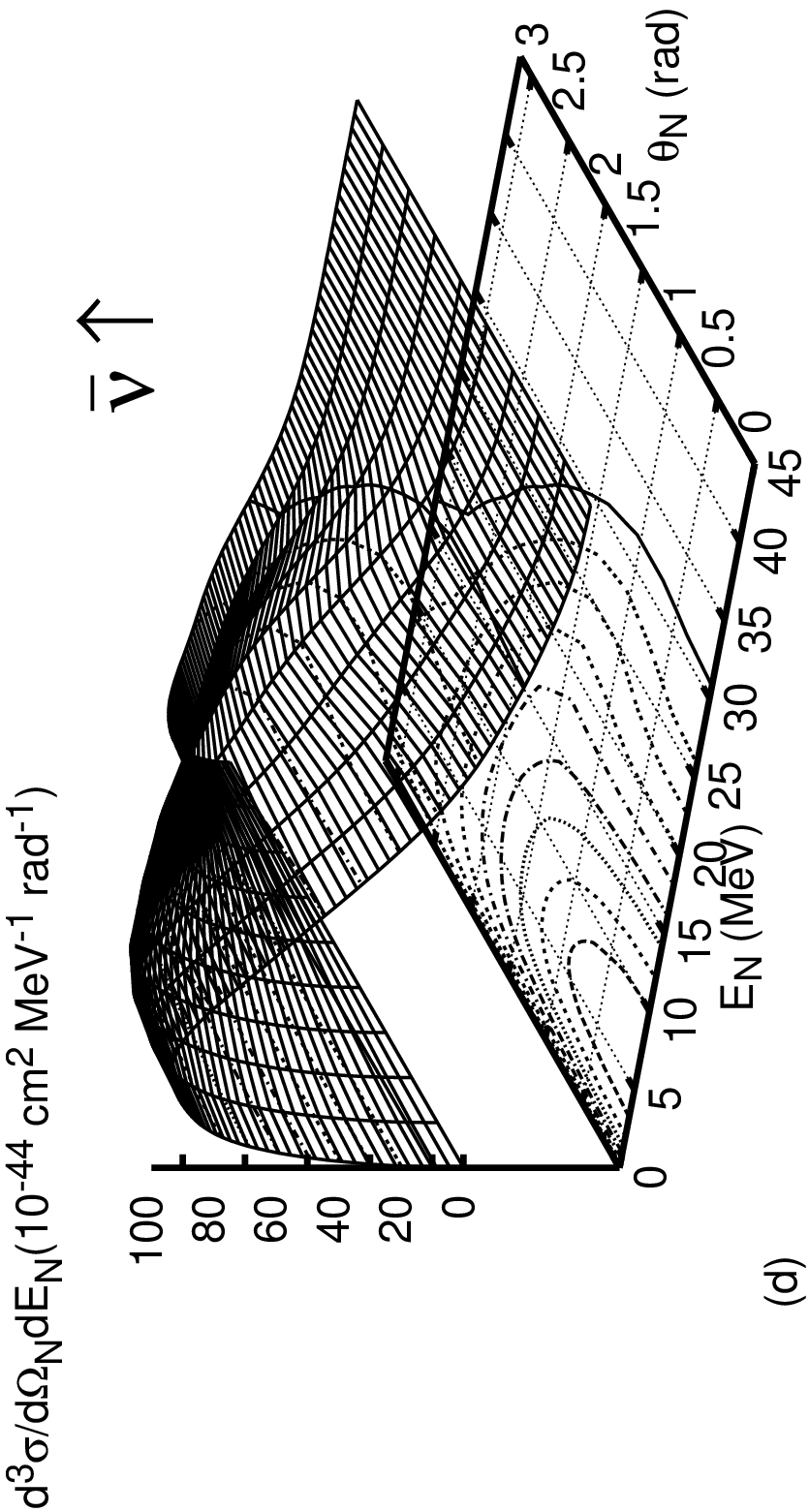"}
\caption{Double differential cross sections for neutrino-induced neutron knockout from $^{208}$Pb.  The upper panels show the neutrino-induced cross sections, the lower panels the cross section for antineutrino reactions. Processes resulting in spin-down nucleon knockout are in the left panels, spin-up cross sections are shown on the right. The incoming lepton energy is 50 MeV.} 
\label{3d}
\end{figure*}

Figure \ref{3d} shows  $\nu$- and $\overline{\nu}$-induced neutron knockout from $^{208}$Pb as a function of the energy of the outgoing nucleon and its scattering angle.  The figure illustrates the polarization asymmetries and the dissimilarities between neutrino- and antineutrino behavior.
The structure of the cross section surfaces nicely shows how neutrino induced $s_N^l=-1$ and antineutrino-induced 
$s_N^l=+1$ cross sections stem from the same $v_T-v_{T'}$ determined lepton behavior  and how the same is true for neutrino induced $s_N^l=+1$ and antineutrino-induced 
$s_N^l=-1$ knockout with the $v_T+v_{T'}$ lepton kinematic factor.  The difference in absolute values between the corresponding curves is due to differences in the hadron responses $h_+{h_+}^*$ and $h_-{h_-}^*$.

\begin{table}
\begin{center}
\begin{tabular}{r|c|c}
\hline
&$h_+{h_+}^*$&$h_-{h_-}^*$ \\
\hline$G_E^2$&$\frac{-k_+k_-}{m_N^2}\delta_{r,m_s}\delta_{r,m'_s}$ & $\frac{-k_+k_-}{m_N^2}\delta_{r,m_s}\delta_{r,m'_s}$\\$G_EG_A$&$\frac{-k_+\sqrt{2}}{m_N}\delta_{r,\frac{1}{2}}\delta_{m_s,\frac{1}{2}}\delta_{m'_s,-\frac{1}{2}}$&$\frac{-k_+\sqrt{2}}{m_N}\delta_{r,-\frac{1}{2}}\delta_{m_s,\frac{1}{2}}\delta_{m'_s,-\frac{1}{2}}$\\&$\frac{+k_-\sqrt{2}}{m_N}\delta_{r,\frac{1}{2}}\delta_{m'_s,\frac{1}{2}}\delta_{m_s,-\frac{1}{2}}$&$\frac{+k_-\sqrt{2}}{m_N}\delta_{r,-\frac{1}{2}}\delta_{m'_s,\frac{1}{2}}\delta_{m_s,-\frac{1}{2}}$\\$G_EG_M\frac{\kappa}{2m_N}$&$\frac{-k_+\sqrt{2}}{m_N}\delta_{r,\frac{1}{2}}\delta_{m_s,\frac{1}{2}}\delta_{m'_s,-\frac{1}{2}}$&$\frac{+k_+\sqrt{2}}{m_N}\delta_{r,-\frac{1}{2}}\delta_{m_s,\frac{1}{2}}\delta_{m'_s,-\frac{1}{2}}$\\&$\frac{+k_-\sqrt{2}}{m_N}\delta_{r,\frac{1}{2}}\delta_{m'_s,\frac{1}{2}}\delta_{m_s,-\frac{1}{2}}$& $\frac{-k_-\sqrt{2}}{m_N}\delta_{r,-\frac{1}{2}}\delta_{m'_s,\frac{1}{2}}\delta_{m_s,-\frac{1}{2}}$\\$G_A^2$&$2\delta_{r,\frac{1}{2}}\delta_{m_s,-\frac{1}{2}}\delta_{m'_s,-\frac{1}{2}}$ & $2\delta_{r,-\frac{1}{2}}\delta_{m_s,\frac{1}{2}}\delta_{m'_s,\frac{1}{2}}$\\$G_M^2\frac{\kappa^2}{4m_N^2}$&$2\delta_{r,\frac{1}{2}}\delta_{m_s,-\frac{1}{2}}\delta_{m'_s,-\frac{1}{2}}$&$2\delta_{r,-\frac{1}{2}}\delta_{m_s,\frac{1}{2}}\delta_{m'_s,\frac{1}{2}}$\\$G_AG_M\frac{\kappa}{2m_N}$&$2\delta_{r,\frac{1}{2}}\delta_{m_s,-\frac{1}{2}}\delta_{m'_s,-\frac{1}{2}}$&$-2\delta_{r,-\frac{1}{2}}\delta_{m_s,\frac{1}{2}}\delta_{m'_s,\frac{1}{2}}$ 
\\
\hline
\end{tabular}
\caption{Spin dependence of various contributions to the transverse  
$h_+{h_+}^*$ and $h_-{h_-}^*$ terms in the hadronic transition density.  
$G_A^2$, $G_E^2$, $G_M^2$,  $G_EG_A$, $G_EG_M$, and $G_AG_M$ denote the axial vector, vector Coulomb, and weak magnetic form factors, and their various interference terms.
The spherical components of the momentum of the outgoing nucleon are  denoted by $k$. Further, $m_s$ and $r$ are the spin components of the bound and outgoing nucleon along the quantization axis determined by $\vec{q}$. }
\end{center}
\label{tabel111}
\end{table}

\section{Summary}

We have made a systematic study of neutrino- and antineutrino-induced nucleon-knockout from $^{16}$O and $^{208}$Pb, focusing on the polarization properties of the ejectile.  The transverse contributions prevailing the cross section, combined with the prominence of backward lepton scattering and forward nucleon knockout, contribute to  a very distinct spin asymmetry signal.

As the ejectile polarization provides a way to distinguish between neutrinos and antineutrinos in neutral current scattering reactions it has some important potential applications.  For mu- and tauneutrinos charged-current reactions are inaccessible for neutrino energies below the lepton-production threshold.  This is the case in some interesting situations.  In a lot of processes of astrophysical interest, neutrino energies are far too low for charged $\nu_{\mu}$ or $\nu_{\tau}$ scattering.  It might however be very interesting to differentiate between neutrinos and antineutrinos \cite{omnis,land,letter}.
In CP-violation \cite{cp,boone} studies involving oscillations to heavy flavor neutrinos, the interplay between  mixing angle, mass-differences, oscillation length and neutrino energies may involve the need  to differentiate between neutrinos and antineutrinos at energies below the lepton-production threshold. The polarization asymmetries then provide a  mechanism to  distinguish between neutrinos and antineutrinos.

Undeniably, the above results are incomplete with respect to the treatment of  final-state interactions and meson-exchange currents.  
However, the cross sections are strongly dominated by forward nucleon knockout    and quasi-elastic processes.
The observed polarization effects are  of a such order of magnitude that the 
major trends can be expected to remain unaffected by final-state interactions and delta- or pion-production processes: (anti)neutrinos induce large asymmetries in ejectile polarizations, with a completely opposite behavior for neutrinos and antineutrinos.  Work on implementing final state interactions in neutrino-induced knockout processes is in progress.

\acknowledgments
N.J.~would like to thank C.~Volpe for interesting discussions.
 The authors are grateful to the Fund for Scientific Research (FWO) Flanders and to the University Research Board (BOF) for financial support.


\begin{references}
\bibitem{jan}J.~Ryckebusch, D.~Debruyne, W.~Van Nespen, and S.~Janssen, Phys.~Rev.~C{\bf 60}, 034604 (1999).
\bibitem{lava} P.~Lava, J.~Ryckebusch, B.~Van Overmeire, and S.~Strauch, nucl-th/0407105.
\bibitem{poltrans}S.~Dieterich, P.~Bartsch, D.~Baumann, J.~Bermuth, K.~Bohinc {\it et al.}, Phys.~Lett.~B500, 47 (2001).
\bibitem{strauch1}S.~Strauch {\it et al.}, Phys.~Rev.~Lett.~{\bf 91}, 052301 (2003)
\bibitem{strauch2}S.~Dieterich {\it et al.}, Phys.~Lett.~{\bf B500}, 47 (2001).
\bibitem{Ed}E.~Kolbe, K.~Langanke, S.~Krewald and F.K.~Thielemann, Nucl.~Phys.~{\bf A540}, 599 (1992).
\bibitem{kim}H.~Kim, J.~Piekarewicz, C.J.~Horowitz, Phys.~Rev.~C {\bf 51}, 2739 (1994).
\bibitem{kol1}J.~Engel, E.~Kolbe, K.~Langanke and P.~Vogel, Phys.~Rev.~C {\bf 54}, 2740 (1996).
\bibitem{kol2}E.~Kolbe, K.~Langanke and P.~Vogel, Nucl.~Phys.~{\bf A652}, 91 (1996).
\bibitem{albe} W.M.~Alberico, M.B.~Barbaro, S.M.~Bilenky, J.A.~Caballero, C.~Giunti, C.~Maieron, E.~Moya de Guerra and  J.M.~Ud\'{\i}as, Nucl.~Phys.~{\bf A623}, 471 (1997). 
\bibitem{sing} S.K.~Singh, N.C.~Mukhopadhyay and E.~Oset, Phys.~Rev.~C {\bf 57}, 2687 (1998).
\bibitem{ikke0} N.~Jachowicz, S.~Rombouts, K.~Heyde, and J.~Ryckebusch,
Phys.~Rev.~C {\bf 59}, 3246 (1999).
\bibitem{fuller} G.~Fuller, W.C.~Haxton, and G.C.~McLaughlin, Phys.~Rev.~D {\bf 59}, 085005 (1999). 
\bibitem{town}A.C.~Hayes and I.S.~Towner, Phys.~Rev.~C {\bf 61}, 044603 (2000).
\bibitem{vol}C.~Volpe, N.~Auerbach, G.~Col\'o, T.~Suzuki, N.~Van Giai, Phys.~Rev.~C {\bf 62}, 015501 (2000).
\bibitem{edw} E.~Kolbe and K.~Langanke Phys.~Rev.~C {\bf 63}, 025802 (2001).
\bibitem{ikke1}N.~Jachowicz, K.~Heyde, J.~Ryckebusch, and S.~Rombouts, Phys.~Rev.~C{\bf 65}, 025501 (2002).
\bibitem{ikke2}N.~Jachowicz, K.~Heyde, and J.~Ryckebusch, Phys.~Rev.~C{\bf 66}, 055501 (2002).
\bibitem{ikke3}N.~Jachowicz and K.~Heyde, Phys.~Rev.~C{\bf 68}, 055502 (2003).
\bibitem{volpenieuw}J.~Engel, G.C.~McLaughlin, C.~Volpe, Phys.~Rev.~D {\bf 67}, 013005 (2003).
\bibitem{edlan}E.~Kolbe, K.~Langanke, G.~Mart\'{\i}nez-Pinedo, and P.~Vogel, J.~Phys.~{\bf G29}, 2569 (2003).
\bibitem{crist} C.~Maieron, M.C.~Martinez, J.A.~Caballero, and J.M.~Ud\'{\i}as, Phys. Rev.~C {\bf 68}, 048501 (2003).
\bibitem{meu} A.~Meucci, C.~Giusti, F.D.~Pacati, Nucl.~Phys.~{\bf A739}, 277-290 (2004). 
\bibitem{nieves}J.~Nieves, J.E.~Amaro, M.~Valverde, nucl-th/0408005.
\bibitem{kub1}M.~Fukugita, Y.~Koyama, and K.~Kubodera, Phys.~Lett.~{\bf B212}, 139 (1988).
\bibitem{kub2}T.~Kuramoto, M.~Fukugita, Y.~Koyama, and K.~Kubodera, Nucl.~Phys.~{\bf A512}, 711 (1990).
\bibitem{kub3}K.~Kubodera and S.~Nozawa, Int.~Journ.~Mod.~Phys.~3, 101 (1994).
\bibitem{Heger}A.~Heger, E.~Kolbe, W.C.~Haxton, K.~Langanke, G.~Mart\'{\i}nez-Pinedo, and S.E.~Woosley, astro-ph/0307546.
\bibitem{letter}N.~Jachowicz, K.~Vantournhout, J.~Ryckebusch, K.~Heyde, Phys. Rev. Lett. {\bf 93}, 082501 (2004).
\bibitem{wso}S.~Krewald, J.~Birkholz, A.~Faessler, and J.~Speth, Phys.~Rev.~Lett.~{\bf 33}, 1385 (1974).
\bibitem{wspb}G.A.~Rinker and J.~Speth, Nucl.~Phys.~A306, 360 (1978).
\bibitem{omnis}R.N.~Boyd and A.St.J.~Murphy, Nucl.~Phys.~{\bf A688}, 386c (2001).
\bibitem{land}C.K.~Hargrove {\it et al.}, Astroparticle Physics {\bf 5}, 183 (1996).
\bibitem{cp}T.~Hattori, T.~Hasuike, S.~Wakaizumi, Phys.~Rev.~D {\bf 65}, 073027 (2002).
\bibitem{boone}E.~Church, I.~Stancu, G.J.~VanDalen, R.A.~Johnson, J.M.~Conrad {\it et al.}, nucl-ex/9706011.
\end{references}
\end{document}